\documentclass[letter,longauth]{aa} 

\usepackage{graphicx}
\usepackage{txfonts}
\usepackage{amssymb,natbib}
\usepackage[breaklinks, colorlinks, citecolor=blue]{hyperref}
\usepackage{xcolor}
\usepackage{hyperref}
\usepackage[]{color}
\usepackage{soul}

\newcommand{\Caii}{Ca\,\textsc{ii}\,HK}
\def\oiipair{[\ion{O}{ii}] $\lambda \lambda$3727,3730}
\def\oii{\hbox{[\ion{O}{ii}]}}
\def\msun {\hbox{M$_\odot$}}
\def\kms{{\hbox{km\,s$^{-1}$}}}
\def\av{${A_{V}}$}
\def\ergcm2s{\ifmmode {\rm\,ergs\,cm^{-2}\,s^{-1}}\else
    ${\rm\,ergs\,cm^{-2}\,s^{-1}}$\fi}
\newcommand{\JWST}{\textit{JWST}}
\newcommand{\HST}{\textit{HST}}
\newcommand{\Spitzer}{\textit{Spitzer}}
\newcommand{\Planck}{\textit{Planck}}

\begin{document}

   \title{Spectroscopy of the Supernova H0pe Host Galaxy at Redshift 1.78}

   \subtitle{}

\author{M.~Polletta\inst{\ref{inst1}}
\and M.~Nonino\inst{\ref{inst2}} 
\and B. Frye\inst{\ref{inst3}}
\and A. Gargiulo\inst{\ref{inst1}}
\and S. Bisogni\inst{\ref{inst1}}
\and N. Garuda\inst{\ref{inst3}}
\and D. Thompson\inst{\ref{inst4}}
\and M. Lehnert\inst{\ref{inst5}}
\and M. Pascale\inst{\ref{inst6}}
\and S.~P. Willner\inst{\ref{inst24}}
\and P. Kamieneski\inst{\ref{inst10}}
\and R. Leimbach\inst{\ref{inst3}}
\and C. Cheng\inst{\ref{inst6b},\ref{inst6c}}
\and D. Coe\inst{\ref{inst7},\ref{inst8},\ref{inst9}}
\and S.~H. Cohen\inst{\ref{inst10}}
\and C.~J. Conselice\inst{\ref{inst11}}
\and L.~Dai\inst{\ref{inst12}}
\and J. Diego\inst{\ref{inst13}}
\and H. Dole\inst{\ref{inst13b}}
\and S.~P. Driver\inst{\ref{inst14}}
\and J.~C.~J. D'Silva\inst{\ref{inst14},\ref{inst15}}
\and A. Fontana\inst{\ref{inst16}}
\and N. Foo\inst{\ref{inst3}}
\and L.~J. Furtak\inst{\ref{inst17}}
\and N.~A. Grogin\inst{\ref{inst7}}
\and K. Harrington\inst{\ref{inst18}}
\and N.~P. Hathi\inst{\ref{inst7}}
\and R.~A. Jansen\inst{\ref{inst10}}
\and P. Kelly\inst{\ref{inst19}}
\and A.~M. Koekemoer\inst{\ref{inst7}}
\and C. Mancini\inst{\ref{inst1}}
\and M.~A. Marshall\inst{\ref{inst20},\ref{inst15}}
\and J.~D.~R. Pierel\inst{\ref{inst7}}
\and N. Pirzkal\inst{\ref{inst7}}
\and A. Robotham\inst{\ref{inst14}}
\and M.~J. Rutkowski\inst{\ref{inst21}}
\and R.~E. Ryan, Jr.\inst{\ref{inst7}}
\and J.~M. Snigula\inst{\ref{inst22}}
\and J. Summers\inst{\ref{inst10}}
\and S. Tompkins\inst{\ref{inst10}}
\and C.~N.~A. Willmer\inst{\ref{inst3}}
\and R.~A. Windhorst\inst{\ref{inst10}}
\and H. Yan\inst{\ref{inst23}}
\and M.~S. Yun\inst{\ref{inst25}}
\and A. Zitrin\inst{\ref{inst17}}
}

 \offprints{M. Polletta\\ \email{maria.polletta@inaf.it}}
\institute{
INAF – Istituto di Astrofisica Spaziale e Fisica Cosmica Milano,  Via A. Corti 12, I-20133 Milano, Italy\label{inst1}
\and INAF-Osservatorio Astronomico di Trieste, Via Bazzoni 2, 34124 Trieste, Italy\label{inst2}
\and Steward Observatory, University of Arizona, 933 N Cherry Ave, Tucson, AZ, 85721-0009, USA\label{inst3}
\and Large Binocular Telescope Observatory, 933 North Cherry Ave, Tucson, AZ 85721, USA\label{inst4}
\and Univ. Lyon, Univ. Lyon1, ENS de Lyon, CNRS, Centre de Recherche Astrophysique de Lyon UMR5574, F-69230 Saint-Genis-Laval, France\label{inst5}
\and Department of Astronomy, University of California, 501 Campbell Hall \#3411, Berkeley, CA 94720, USA\label{inst6}
\and Center for Astrophysics \textbar\ Harvard \& Smithsonian, 60 Garden Street, Cambridge, MA 02138, USA\label{inst24}
\and School of Earth and Space Exploration, Arizona State University, Tempe, AZ 85287-1404, USA\label{inst10}
\and Chinese Academy of Sciences South America Center for Astronomy, National Astronomical Observatories, CAS, Beijing 100101, China\label{inst6b}
\and CAS Key Laboratory of Optical Astronomy, National Astronomical Observatories, Chinese Academy of Sciences, Beijing 100101, China\label{inst6c} 
\and Space Telescope Science Institute, 3700 San Martin Drive, Baltimore, MD 21218, USA\label{inst7}
\and Association of Universities for Research in Astronomy (AURA) for the European Space Agency (ESA), STScI, Baltimore, MD 21218, USA\label{inst8}
\and Center for Astrophysical Sciences, Department of Physics and Astronomy, The Johns Hopkins University, 3400 N Charles St. Baltimore, MD 21218, USA\label{inst9}
\and Jodrell Bank Centre for Astrophysics, Alan Turing Building, University of Manchester, Oxford Road, Manchester M13 9PL, UK\label{inst11}
\and Department of Physics, University of California, 366 Physics North MC 7300, Berkeley, C. 94720, USA\label{inst12}
\and FCA, Instituto de Fisica de Cantabria (UC-CSIC), Av.  de Los Castros s/n, E-39005 Santander, Spain\label{inst13}
\and Université Paris-Saclay, CNRS, Institut d'Astrophysique Spatiale, 91405, Orsay, France\label{inst13b}
\and International Centre for Radio Astronomy Research (ICRAR) and the International Space Centre (ISC), The University of Western Australia, M468, 35 Stirling Highway, Crawley\label{inst14}
\and ARC Centre of Excellence for All Sky Astrophysics in 3 Dimensions (ASTRO 3D), Australia\label{inst15}
\and INAF Osservatorio Astronomico di Roma, Via Frascati 33, I-00078 Monteporzio Catone, Rome, Italy\label{inst16}
\and Physics Department, Ben-Gurion University of the Negev, P. O. Box 653, Be’er-Sheva, 8410501, Israel\label{inst17}
\and European Southern Observatory, Alonso de C\'ordova 3107, Vitacura, Casilla 19001, Santiago de Chile, Chile\label{inst18}
\and School of Physics and Astronomy,  University of Minnesota, 116 Church Street SE, Minneapolis, MN 55455, USA\label{inst19}
\and National Research Council of Canada, Herzberg Astronomy \& Astrophysics Research Centre, 5071 West Saanich Road, Victoria, BC V9E 2E7, Canada\label{inst20}
\and Minnesota State University-Mankato,  Telescope Science Institute, TN141, Mankato MN 56001, USA\label{inst21}
\and Max-Planck-Institut für extraterrestrische Physi, Garching, 85741, Germany\label{inst22}
\and Department of Physics and Astronomy, University of Missouri, Columbia, MO 65211, USA\label{inst23}
\and Department of Astronomy, University of Massachusetts at Amherst, Amherst, MA 01003, USA\label{inst25}
}

\authorrunning{Polletta et~al.}

\date{Received 22 May 2023 / Accepted 20 June 2023}

  \abstract
   {Supernova (SN) H0pe was discovered as a new
   transient in James Webb Space Telescope (\JWST) NIRCam images of the galaxy cluster \object{PLCK\,G165.7$+$67.0} taken as part of the ``Prime Extragalactic Areas for Reionization and Lensing Science'' (PEARLS) \JWST\ GTO program (\# 1176) on 2023 March 30~\citep[AstroNote 2023-96; ][]{frye23_astronote}. 
   The transient is a compact source associated with a background galaxy that is stretched and triply-imaged by the cluster's strong gravitational lensing. This paper reports spectra in the 950--1370\,nm  observer frame of two of the galaxy's images obtained with Large Binocular Telescope (LBT) Utility Camera in the Infrared (LUCI) in longslit mode  two weeks after the \JWST\ observations. The  individual and average spectra show the \oiipair\ doublet and the Balmer and 4000\,\AA\ breaks at redshift $z{=}$1.783$\pm$0.002. 
   The Code Investigating GAlaxy Emission (CIGALE) best-fit model of the spectral energy distribution indicates that SN H0pe's host galaxy is massive (${M}_{\rm star}{\simeq}6{\times}10^{10}$\,\msun\ after correcting for a magnification factor $\mu\sim7$) with a predominant intermediate age ($\sim$2\,Gyr) stellar population, moderate extinction, and a magnification-corrected star formation rate $\simeq$13\,\msun\,yr$^{-1}$, consistent with being below the main sequence of star formation.  These properties suggest that H0pe might be a type Ia SN. 
   Additional observations of SN H0pe and its host recently carried out with \JWST\ (\JWST-DD-4446; PI: B.\ Frye)  will be able to both determine the SN classification and confirm its association with the galaxy analyzed in this work. 
   }

   \maketitle
%

\section{Introduction}

The James Webb Space Telescope~\citep[\JWST;][]{Gardner2006, Rieke2005, Beichman2012, Windhorst2008}, with its unprecedented sensitivity and spatial resolution at infrared wavelengths, can be used as a time machine to capture the light even from single stars early in the history of the Universe from their birth in a dusty cradle to their spectacular death in the form of supernovae (SNe) \citep{welch22,diego23,vanzella23,meena23,kelly23}. 

\begin{figure}[h!]
\centering 
\includegraphics[width=0.45\textwidth]{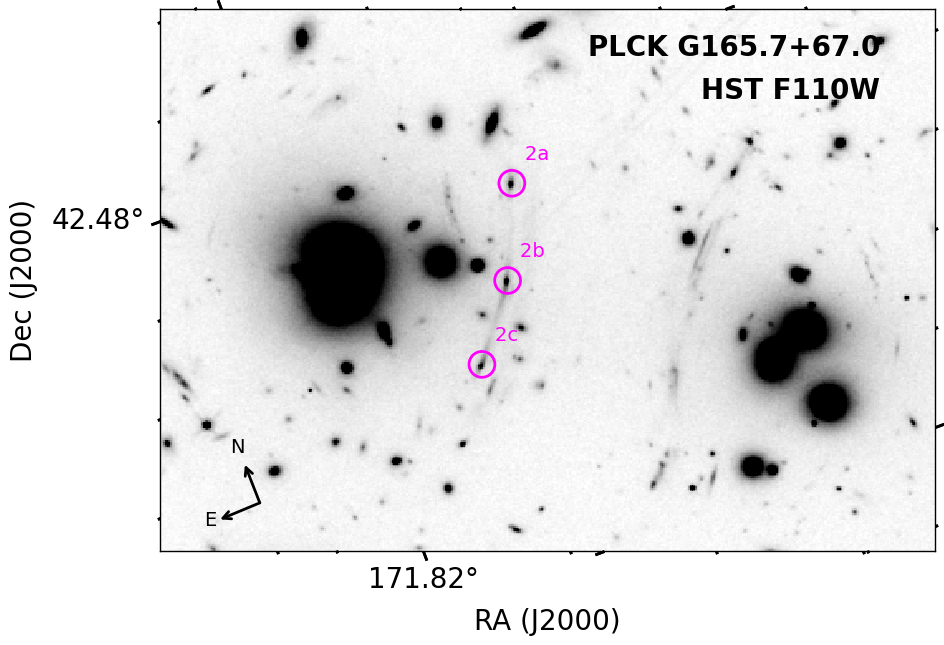}
\caption{\small Archival \textit{\HST}/WFC3 F110W negative image of the PLCK\,G165.7$+$67.0 field (1\arcmin$\times$0.7\arcmin). The triply-imaged arcs 2a, 2b, and 2c of the SN H0pe host galaxy are highlighted with 1\arcsec-radius magenta circles \citep[credits: ][]{frye19}. The highlighted regions are part of an arc produced by strong lensing from the foreground galaxy cluster.}
\label{fig:g165_image} 
\end{figure}

Recently,~\citet{frye23_astronote} announced the discovery of a SN in a galaxy at a photometric redshift of $\sim$1.8 in \JWST/NIRCam multi-band observations obtained as part of the ``Prime Extragalactic Areas for Reionization and Lensing Science'' (``PEARLS'') GTO Program~\citep[\# 1176, PI: Windhorst; ][]{windhorst23} in the PLCK\,G165.7$+$67.0 (G165 hereinafter) field.  
G165 is a massive ($M_{\rm 600 kpc} = 2.4 \times 10^{14}$\,\msun) galaxy cluster at $z{=}0.35$ \citep{canameras18,frye19,pascale22}. The cluster strongly lenses numerous background galaxies including bright sub-millimeter (sub-mm) galaxies at $z{\ga}$2 \citep{harrington16}, and indeed G165 was discovered by the \Planck\ telescope because of its bright sub-mm emission \citep{planck15,planck16}. Comparison of the new \JWST/NIRCam images with psf-matched archival images from Wide Field Camera 3 (WFC3) on the \textit{Hubble Space Telescope}~\citep[\HST;][]{Miley2004,Zirm2005} revealed a new, pointlike source associated with a triply-imaged galaxy.
The magnification factor of the lensed galaxy is estimated to be $\mu\sim3-16$ across the three images, designated as 2a, 2b, and 2c \citep[Fig.~\ref{fig:g165_image};][Kamieneski, priv. comm.]{frye19,pascale22}. The new source, visible next to all three of the lensed galaxy images (see Fig.~\ref{fig:rgb_img}), was named ``\object{SN H0pe}'' because it offers the potential of a new, independent measurement of the Hubble--Lema\^itre constant $H_0$\null. 

The apparent SN discovery prompted follow-up spectroscopic observations with the goal of measuring the spectroscopic redshift of the host galaxy. The redshift was needed to plan effective observations to determine the redshift and type of the SN\null. Because the SN light fades with time, these observations had to be carried out as soon as possible after the SN's discovery. Through a concerted effort of the Large Binocular Telescope \citep[LBT;][]{hill12} consortium, longslit spectroscopic observations with the LBT Utility Camera in the Infrared~\citep[LUCI; ][]{seifert03,seifert10,ageorges10,buschkamp12} were quickly planned, carried out, reduced, and analyzed. 

This Letter presents the results of the LUCI observations:  the first spectroscopic measurement and characterization of the SN H0pe host galaxy. The study of even a single SN host galaxy is important to learn about SN progenitors, SN host demographics, and to quantify age, metallicity, and dust extinction effects on SN measurements. This is especially relevant at high redshift as some evolutionary trends could mimic cosmological effects. 

Additional spectroscopic and multi-epoch imaging observations were carried out with \JWST\ after the LBT observations (Program \JWST-DD-4446; PI: B.\ Frye). These observations, which will be presented in a future work, will help to establish the SN type, verify its association with the triply-imaged galaxy, and measure $H_0$.
Throughout this study, we adopt a \citet{chabrier03} initial mass function (IMF) and a flat $\Lambda$ cold dark matter (CDM) model with cosmological parameters from  the Planck 2018 release \citep[i.e., $\Omega_{\Lambda}$\,=\,0.685; $\Omega_{\mathrm{m}}$\,=\,0.315; H$_{\mathrm{0}}$\,=\,67.4\,\kms\,Mpc$^{-1}$; ][]{planck_cosmo18}.  All colors and magnitudes quoted in this paper are expressed in the AB system~\citep{OkeGunn1983}.

\section{Large Binocular Telescope spectroscopic observations}\label{sec:spec_obs}

\begin{figure}[h!]
\centering 
\includegraphics[height=7cm]{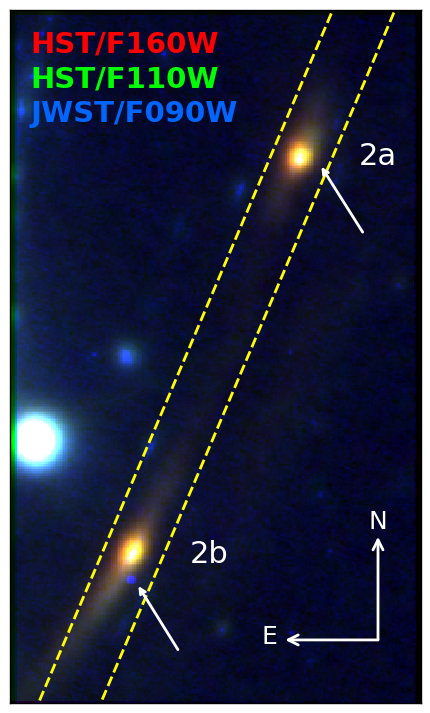}
\includegraphics[height=7cm]{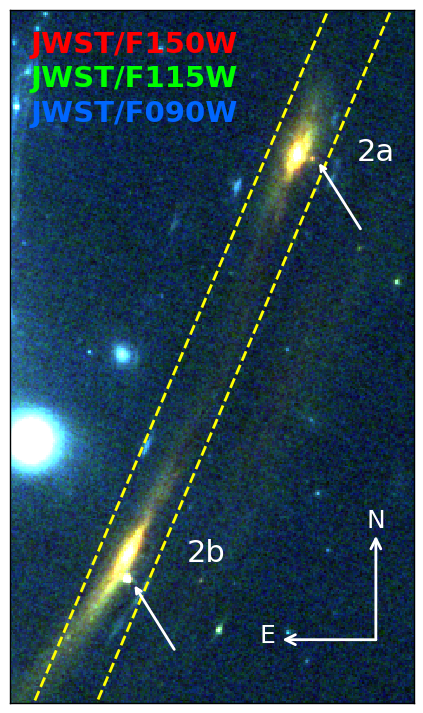}
\caption{\small Multi-band 7\arcsec$\times$12\arcsec\ images of SN H0PE and its host galaxy. \textit{Left panel:} HST images taken in 2015--2016 (red: HST/F160W, green: HST/F110W, Cy23, GO-14223, PI: Frye) and
\JWST\ image (blue: F090W) taken on 2023 March 30, about two weeks before the LBT spectroscopic observation reported here. \textit{Right panel:}
\JWST\ images taken on 2023 March 30 (red: \JWST/F150W, green: \JWST/F115W, and blue:
\JWST/F090W). \JWST\ images are from the PEARLS GTO program and are adapted from \url{https://www.as.arizona.edu/bright-and-shiny-result-pearls}.  The two SN H0pe host-galaxy images 2a and 2b are labeled, and the SN positions are marked with white arrows. Dashed yellow lines mark the slit edges at PA\,=\,157\degr.}
\label{fig:rgb_img} 
\end{figure}

Longslit spectroscopic observations of the SN host were obtained with LUCI on the LBT (Program INAF-2023A-777; PI: M.\  Nonino). Observations were carried out on UT 2023 April 16 by the LBTB team with the support of the University of Arizona LBT staff. Observations used the N1.8 camera (0\farcs249/pixel) and the second order of the G200 grating with the zJspec filter resulting in wavelength coverage from 950 to 1370\,nm in the observer frame. The slit width was 1\arcsec, which results in a spectral resolution R$\sim$1200, sufficient to identify the 400\,nm break, strong emission lines, and absorption features at the expected source redshift. The main goal of these observations was to obtain a spectroscopic redshift and the optical rest-frame spectrum of two images of the galaxy associated with SN H0pe, sources \object{2a} ($\alpha_{\rm 2000}$\,=\,11:27:15.33; $\delta_{\rm 2000}$\,=\,42:28:41.07), and \object{2b} ($\alpha_{\rm 2000}$\,=\,11:27:15.60; $\delta_{\rm 2000}$\,=\,42:28:34.13) \citep{frye19}. To center the 1\arcsec$\times$200\arcsec\ longslit on source 2b, the brighter of the two target images, the telescope was blindly offset by $\sim$5\arcsec\ from a bright source. The slit had a position angle (PA) of
157\degr\ in order to include also image 2a of the SN host.  Fig.~\ref{fig:rgb_img} shows a multi-band \JWST\ image from the PEARLS Program of the two host images with  the longslit position superimposed. The respective SN images were also in the slit but probably contributed $<$8\% of the observed light  ($m$(F160W)$_{\rm 2b}{\simeq}$21.12 {versus} $m$(F150W)$_{\rm SNb}{\simeq}$23.91; \citealt{pascale22,frye23_astronote}.) Eight additional objects, of which six are within the \JWST/NIRCam field-of-view, were serendipitously detected in the slit. These other objects will be discussed in a future work.

The observations were carried out in binocular mode and divided into 36 300\,s exposures per arm, resulting into a total exposure time of 6\,hours. The telescope was ``nodded'' along the direction of the slit with 4\arcsec\ offsets to obtain good sky subtraction during the data reduction. A star was observed at the end of the observations for flux calibration and to correct for atmospheric absorption. Flat-field exposures were taken on the same day as the observations. The observing conditions were good with a clear sky, an average airmass of 1.03, and seeing 1\arcsec. 

Data reduction was carried out by the Italian LBT support team using the Spectroscopic Interactive Pipeline and Graphical Interface
(SIPGI) pipeline developed at INAF IASF-Milan \citep{gargiulo22}.  This included flat-fielding, dark subtraction, sky subtraction, correction for bad pixels, cosmic rays, and distortions, and wavelength and flux calibration. From each scientific exposure, the background level was first removed by subtracting the subsequent dithered frame in the observing sequence. Possible background residuals were then estimated and removed by fitting all the pixels on the sky along the spatial direction of each column with a polynomial. Wavelength calibration was carried out using OH sky lines and reaches an accuracy of 0.25\,\AA. For each source detected in the final flux-calibrated 2-d spectrum, the flux calibrated 1-d spectrum was extracted, and the associated sky and noise spectra were produced. 

\section{Spectroscopic redshift measurement}\label{sec:spectra}

\begin{figure*}[ht!]
  \begin{center}
 \includegraphics[width=0.8\textwidth]{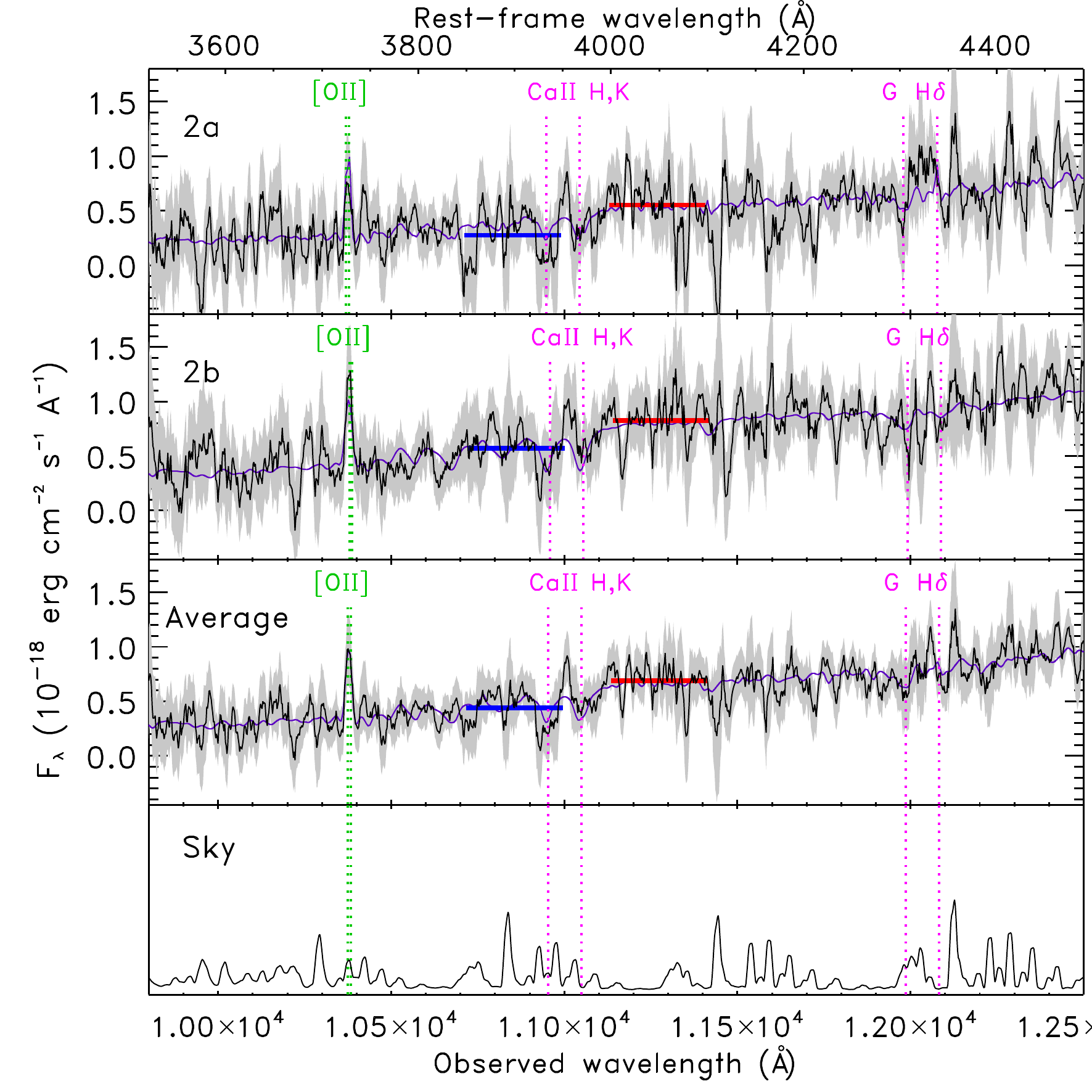}
  \end{center}
 \caption{\small Smoothed ($\Delta\lambda{=}17.2$\,\AA) and cropped LBT/LUCI spectra (black solid line) of the SN H0pe host from image 2a ({\it top panel}), 2b ({\it second panel}), the averaged spectrum ({\it third panel}), and the sky ({\it bottom panel}). The grey shaded area shows the 1$\sigma$ noise.  The main
spectral features are labeled with green (for emission) and magenta (for absorption)
vertical dotted lines. The purple line represents the Bagpipes best-fit model.  The average flux densities in the rest-frame wavelength intervals 385--394\,nm and 400--410\,nm, used to compute the D4000$_n$ index, are shown with blue and red horizontal lines, respectively.}
   \label{fig:lbt_spectra}
\end{figure*}

The LBT spectra of 2a and 2b are shown in Fig.~\ref{fig:lbt_spectra} along with their average, and the sky spectrum.
We measured the redshift of each spectrum and of the averaged one using the
EZ code \citep{garilli10}.  EZ derives the redshift through a weighted
cross-correlation between the observed spectrum over a specific wavelength
range (980--1350\,nm in this case) and a library of galaxy spectral
templates.  EZ takes into account the presence of strong emission features
as well as of the continuum shape in the redshift determination.  EZ
selected as best-fit template a red elliptical at ${\langle}z_{\rm
spec}{\rangle}{=}$1.783 ($z{=}$1.782, 1.784, and 1.783 for 2a, 2b, and the
averaged spectrum, respectively).  Because EZ uses a library of templates
rather than actual fits, we derived the best spectral model using the
Bagpipes code \citep{carnall21}.  The redshifts were fixed at the measured
values, and other parameters included a double-exponential star-formation
history (SFH), the stellar population models of \citet{bruzual03}, and the
\citet{calzetti00} attenuation law.  The Bagpipes best fit-models at the EZ
spectroscopic redshifts, shown in Fig.~\ref{fig:lbt_spectra}, are consistent
with the spectra having \Caii\ K and H absorption lines (at rest
3934.8\,\AA\ and 3969.6\,\AA), the \oiipair\ doublet, and the Balmer
(3646\,\AA) and 4000\,\AA\ breaks.  The signal-to-noise ratios (SNRs) of the
\oii, K and H lines, computed as the difference between the line peak and
the interpolated continuum at the line wavelength divided by the rms of the
spectrum in a 200\,\AA\ region centered on each line, are, respectively,
3.5, 0.6, and 0.7.  The measured Balmer and 4000\,\AA\ breaks\footnote{The
4000\,\AA\ break is quantified through the D4000$_n$ index \citep{balogh99}. 
This was computed as the ratio between the average flux density in the
wavelength intervals 400--410\,nm and 385--395\,nm.  The Balmer break was
computed as the ratio between the average flux density in the wavelength
intervals 380--395\,nm and 350--365\,nm.} are, respectively, 1.48$\pm$0.25
and 1.57$\pm$0.25.  While the only spectral features significantly detected
individually are the \oii\ doublet and the breaks in the continuum, all
features contribute to the correlation signal.  The width of \oii\ doublet
constrains the spectroscopic redshift to be between 1.781 and 1.785,
consistent with the EZ measurement, and we therefore report
$z{=}$1.783$\pm$0.002 as a best estimate and conservative uncertainty.  This
spectroscopic redshift is an important factor for obtaining a robust lens
model of the triply-imaged system and enabling precision cosmology with SN
H0pe~\citep{johnson16}.

\section{Properties of the SN host galaxy}\label{sec:host_properties}

To characterize the SN host, we analyzed its spectral energy distribution
(SED)\null.  For this study, we consider only source 2b, for which
multi-band photometric measurements are available \citep[see ID~3 in Table 2
of][]{pascale22}.  The data include LBT/LBC ($g$ and $i$), \HST/WFC3 (F110W,
F160W), LBT/LUCI ($K$), and \Spitzer/IRAC (3.6, 4.5\,$\mu$m).  Based on a
preliminary analysis of the \JWST/NIRCam imaging, these measurements are
affected by a magnification factor $\mu{\approx}7.4^{+8.9}_{-3.7}$ (P. 
Kamieneski, priv.  comm.).
\begin{figure}[ht!]
  \begin{center}
 \includegraphics[width=0.49\textwidth]{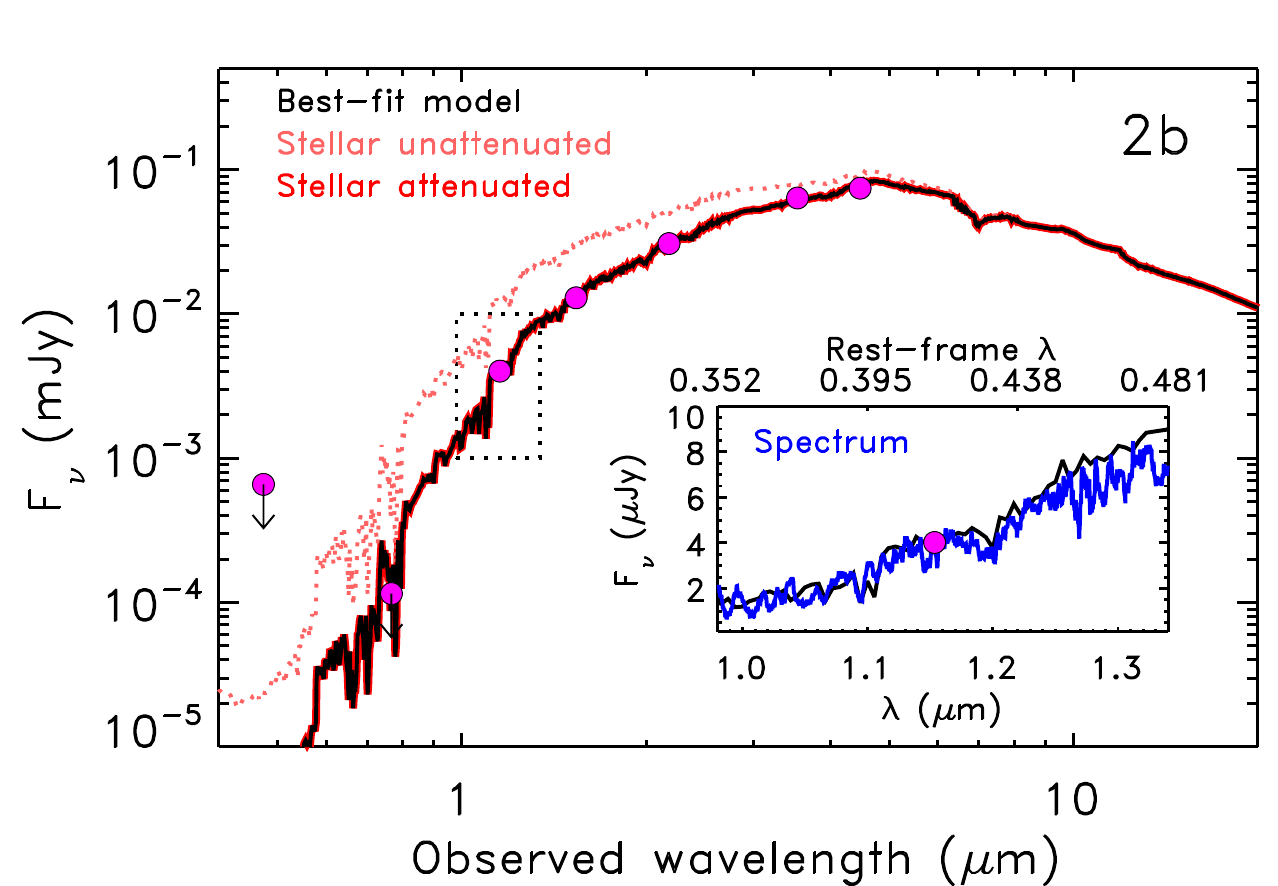}
  \end{center}
 \caption{\small Spectral energy distributions of the SN H0pe host image
2b.  Magenta filled circles show the observed photometry, not corrected for
magnification, and arrows represent 5$\sigma$ upper limits.  The CIGALE
best-fit model is shown with a black solid line, the light red dotted line
shows the stellar light before dust attenuation, and the red solid line
(coincident with the black line) shows the attenuated stellar light.  The
inset shows the smoothed LBT 1-d spectra (blue solid line), the best-fit
SED (black line), and the F110W photometric data point (magenta filled circle)
in the ranges shown by the dotted rectangles in the main panel.}
   \label{fig:cigale_sed}
\end{figure}
We modeled the SED of source 2b with the Code Investigating GAlaxy
Emission~\citep[CIGALE;][]{boquien19}.  The code provides constraints on the
age of the stellar components and estimates of dust extinction, stellar
mass, and SFR\null.  We fixed the redshift at the EZ value obtained from the
spectrum and adopted a model with a delayed star formation history (SFH),
a~\citet{chabrier03} initial mass function (IMF), the stellar population
models of \citet{bruzual03}, solar metallicity ({$Z$}\,=\,0.02), and the
\citet{calzetti00} attenuation law.  The best fit was determined as the
template with the lowest $\chi^2$, and the best-fit parameters and
associated uncertainties are the likelihood-weighted means and standard
deviations, respectively.  The results are consistent with the spectral fit
from Bagpipes.  The derived stellar age is 2.5$\pm$0.5\,Gyr, and the
magnified stellar mass and SFR (i.e., as calculated without correcting for
magnification) are, respectively, log($\mu M_{\rm
star}$/\msun)\,=\,11.6$\pm$0.1 and $\mu$SFR$<$130\,\msun\,yr$^{-1}$, the
latter only an upper limit.  The fit requires a significant amount of dust
extinction, \av\,=\,1.2$\pm$0.4.  Fig.~\ref{fig:cigale_sed} shows the model,
the photometry, and the LBT flux-calibrated 1-d spectra.  There is good
agreement between the 1-d spectrum and the SED, demonstrating the goodness
of the spectral flux calibration.  In summary, the CIGALE model indicates
that the SN host is a dust-obscured massive galaxy of intermediate
age.

An independent measurement of the galaxy SFR can be obtained from the
luminosity of the \oii\ doublet, which is produced by ionized gas in
star-forming regions.  (Neither the spectrum nor the photometry show any
evidence of an active galactic nucleus, and no pointlike nucleus is seen in
the \HST\ images.) The \oii\ flux obtained by fitting the doublet with a
Gaussian profile in the average spectrum is
(1.4$\pm$0.2)$\times$10$^{-17}$\ergcm2s.  The SFR--L(\oii) relation depends
on oxygen abundance \citep[eq.\ 5 of][]{zhuang19}.  For the SN host galaxy,
we adopted log(O/H)$+$12\,=\,9.02, the value typical of massive galaxies as
the CIGALE fit shows the SN host to be.\footnote{For this abundance,
SFR/(\msun\,yr$^{-1}$)\,=\,7.74${\times}10^{-42}L_{\oii}$/(erg\,s$^{-1}$)
\citep{zhuang19}.} This gives magnified $\mu \rm
SFR=(2.5\pm0.4)$\,\msun\,yr$^{-1}$ for the average spectrum.  If the \oii\
lines suffer extinction 3.95\,mag,\footnote{The \citet{calzetti00} reddening
curve gives $A_{3728\AA}$\,=\,1.45$\times$\av, but in addition, star-forming
gas clouds have higher extinction than the stellar population
\citep{calzetti97}, $E_{\rm gas}(B{-}V){=}E_{\rm
star}(B{-}V)/(0.44{\pm}0.03)$.} and the magnification factor
$\mu{\approx}$7 (Kamieneski, priv.  comm.), the magnification- and
dust-corrected SFR$\simeq$(13$\pm$2)\,\msun\,yr$^{-1}$.  The true SFR might
be smaller if diffuse stellar sources such as hot post-asymptotic giant
branch stars contribute to \oii\ \citep[e.g.,][]{belfiore16}.

The  main-sequence SFR for a galaxy of the same mass (corrected for
$\mu{\approx}$7) and redshift is 82$^{+48}_{-31}$\,\msun\,yr$^{-1}$, where
the uncertainty is derived from the 0.2\,dex scatter \citep{speagle14}.  The
\oii\ SFR is therefore below the main sequence value, implying that this
system might be transitioning to or already in a passive phase.
The galaxy spectrum (Fig.~\ref{fig:lbt_spectra}) exhibits a 4000\,\AA\
break, which is developed by a passively evolving stellar population at
500\,Myr~\citep{bica94} and strengthens  with age \citep{kauffmann03}.  The
D4000$_n$ value in the average spectrum is 1.6$\pm$0.3, consistent with the
value D4000$_n{\simeq}$1.55 usually adopted to separate star-forming and
passive galaxies  \citep[e.g.,][]{gargiulo17,haines17}.  Thus D4000$_n$
agrees with the CIGALE age and with this system being on its way to becoming
a passive galaxy.

\section{Discussion and perspectives}

Discovery of distant SNe and study of their hosts are important for
precision cosmology because the SN properties (i.e., luminosity at maximum
and rate of decline) depend on host properties
\citep{hamuy95,sullivan06,williams20}, and any systematic trends of the
latter with redshift could mimic a cosmological effect
\citep{williams03,sullivan03,combes04}.  Examples include redshift-dependent
reddening laws \citep{totani99} or element abundance ratios of the
progenitor stars \citep{hoflich20,drell00}.  If the association of SN H0pe
with the triply-imaged galaxies 2a, 2b, and 2c is confirmed, this SN would
be among the most distant known SNe.  Indeed, to our knowledge and without
considering the super luminous SNe~\citep{khetan23}, today there are only
four known SNe at $z{\geq}$1.8 \citep[i.e., 1.80${<}z{<}$2.26;
][]{jones13,rodney15,rubin18}, and SN H0pe is the only one with \JWST\ data. 
This system thus offers an excellent opportunity to study in detail a
high-$z$ SN host, and the SN itself.

The SN H0pe host-galaxy SED indicate that this system, after correcting for
a magnification factor $\mu=7$, is a massive ($M_{\rm
star}\sim(6.0{\pm}0.8)\times10^{10}$\,\msun) galaxy of intermediate age
(2.5$\pm$0.5\,Gyr) with significant extinction and low star-formation rate
($\rm SFR\simeq(13\pm2)$\,\msun\,yr$^{-1}$), implying that it might be on
its way to becoming or be quiescent \citep{speagle14}.  This finding
suggests that H0pe might be a type Ia SN, as these occur in all type of
galaxies, whereas core-collapse SNe are only found in star forming
galaxies~\citep{kelly12}.  Since only one third of SNe Ia is found in
early-type systems, locally and up to intermediate
redshifts~\citep{farrah02}, it is quite exceptional to have caught this SN
in such a galaxy.  Furthermore, SNe are more difficult to observe in
galaxies that suffer from dust obscuration.  Again, SN Hope beats the odds
because of the significant extinction derived for its host.  The host red
color and extinction are unusual among both main-sequence star-forming
galaxies \citep{pacifici23} and quiescent galaxies \citep{deugenio21}.  A
possible explanation is that the obscuring dust might be in the foreground,
perhaps associated with the G165 cluster as previously suggested by
\citet{pascale22}, but this hypothesis seems unlikely as there is no sign of
an intervening galaxy.  Intergalactic dust is also implausible because it
would not cause reddening \citep{aguirre00}.  The amount of dust extinction,
the level of star-formation activity, and the lensing magnification factor
in this galaxy are quite uncertain, and more study is needed to pin down the
values.  However, the overall characterization of the galaxy as massive and
nearing or at quiescence should be secure.  Aside from its intrinsic
interest, characterizing the dust properties of the SN H0pe host is
necessary to correctly analyze the SN measurements.

Analysis of the \JWST\ data from the PEARLS program and the more recent DD
observations will yield an improved SED measurement of the triply-imaged
host and a refined lens model.  These will provide better constraints on the
system star-formation activity level and stellar age.  In addition, it will
be possible to investigate whether the SN suffers dust extinction and how
this might affect the SN measurements.  The \JWST\ data will also determine
the SN classification, test its association with the triply-imaged galaxy,
measure time delays of the three lensed images, and provide a precise
measurement of $H_0$ \citep[see e.g.,][]{kelly23}.

\begin{acknowledgements}
We kindly thank the referee for promptly reviewing the manuscript.
M.P. kindly thanks M. Fumana, and P. Franzetti for their assistance in installing and using EZ, F. Cusano for his prompt help with the LBT data retrieval, and D. Burgarella for his support with CIGALE.
B.L.F.~obtained student support through a Faculty Challenge Grant for Increasing Access to Undergraduate Research, and the Arthur L. and Lee G. Herbst Endowment for Innovation and the Science Dean’s Innovation and Education Fund, both obtained at the University of Arizona.
AZ and LJF acknowledge support by Grant No. 2020750 from the United States-Israel Binational Science Foundation (BSF) and Grant No. 2109066 from the United States National Science Foundation (NSF), and by the Ministry of Science \& Technology, Israel.
The LBT is an international collaboration among institutions in the United States, Italy and Germany. LBT Corporation partners are: The University of Arizona on behalf of the Arizona university system; Istituto Nazionale di Astrofisica, Italy; LBT Beteiligungsgesellschaft, Germany, representing the Max-Planck Society, the Astrophysical Institute Potsdam, and Heidelberg University; The Ohio State University, and The Research Corporation, on behalf of The University of Notre Dame, University of Minnesota, and University of Virginia.
We acknowledge the support from the LBT-Italian Coordination Facility for the execution of observations, data distribution and reduction.
%
This research is based on observations made with the NASA/ESA Hubble Space
Telescope obtained from the Space Telescope Science Institute, which is
operated by the Association of Universities for Research in Astronomy, Inc.,
under NASA contract NAS 5–26555.  These observations are associated with
program 14223.
This work is based on observations made with the NASA/ESA/CSA James Webb Space
Telescope. The data were obtained from the Mikulski Archive for Space
Telescopes at the Space Telescope Science Institute, which is operated by the
Association of Universities for Research in Astronomy, Inc., under NASA
contract NAS 5-03127 for \JWST. These observations are associated with
\JWST\ program \# 1176.
{\it Software:} This research made use of astropy, a community developed core Python package
for astronomy~\citep{astropy}, APLpy, an open-source plotting package for
Python~\citep{aplpy}, the IDL Astronomy Library~\citep{Landsman1993}; SIPGI~\citep{gargiulo22}; Bagpipes~\citep{carnall21}; EZ~\citep{garilli10}; and CIGALE~\citep{boquien19}.

\end{acknowledgements}

\end{document}